\documentclass[twocolumn,10pt, conference]{IEEEtran}
\usepackage{amsmath,amssymb,amsfonts}
\usepackage{algorithmic}
\usepackage{graphicx}
\usepackage{textcomp}
\usepackage{xcolor}
\usepackage{multirow}
\usepackage[T1]{fontenc} 
\usepackage{booktabs}
\usepackage{array}
\usepackage[caption=false]{subfig}
\usepackage{amssymb}
\usepackage{amsmath}
\usepackage{graphicx}
\usepackage{wrapfig}
\graphicspath{ {./images/} }
\usepackage{dcolumn}
\usepackage{hyperref}
\usepackage[labelfont=bf]{caption}
\usepackage{float}

\begin{document}

\title{Machine Learning (ML)-assisted Beam Management in millimeter (mm)Wave Distributed Multiple Input Multiple Output (D-MIMO) systems} 


\author{
  \IEEEauthorblockN{ Karthik R M}
\IEEEauthorblockA{\textit{Senior Researcher} \\
\textit{Ericsson Research}\\
Chennai, India \\
karthik.r.m@ericsson.com}
\and
\IEEEauthorblockN{Dhiraj Nagaraja Hegde}
\IEEEauthorblockA{\textit{Senior Researcher} \\
\textit{Ericsson Research}\\
Chennai, India  \\
dhiraj.nagaraja.hegde@ericsson.com}
\and
\IEEEauthorblockN{Muris Sarajlic}
\IEEEauthorblockA{\textit{Senior Researcher} \\
\textit{Ericsson Research}\\
Lund, Sweden \\
muris.sarajlic@ericsson.com}
\and
\IEEEauthorblockN{Abhishek Sarkar}
\IEEEauthorblockA{\textit{Senior Researcher} \\
\textit{Ericsson Research}\\
Chennai, India \\
abhishek.sarkar@ericsson.com }
}

\maketitle

\begin{abstract}
Beam management (BM) protocols are critical for establishing and maintaining connectivity between network radio nodes and User Equipments (UEs). In Distributed Multiple Input Multiple Output systems (D-MIMO), a number of access points (APs), coordinated by a central processing unit (CPU), serves a number of UEs. At mmWave frequencies, the problem of finding the best AP and beam to serve the UEs is challenging due to a large number of beams that need to be sounded with Downlink (DL) reference signals. The objective of this paper is to investigate whether the best AP/beam can be reliably inferred from sounding only a small subset of beams and leveraging AI/ML for inference of best beam/AP. We use Random Forest (RF), MissForest (MF) and conditional Generative Adversarial Networks (c-GAN) for demonstrating the performance benefits of inference. 
\end{abstract}

  
\section{Introduction and System Model}
\label{sec:intro}
Massive multiple input multiple output (mMIMO) is one of the enabling technologies for $5G$ and $6G$ systems where a large number of antenna elements provide additional degrees of freedom to increase the throughput and provide considerable beamforming gains for improving the coverage. Beamforming concentrates the signal energy in a small angular space.
The beam management (BM) is defined as the process of acquiring and maintaining a set of
beams, which are originated at the gNB and/or the UE and can be used for downlink (DL) and
uplink (UL) transmission and reception. 
BM collectively encompasses initial beam alignment, monitoring,
and tracking, as well as recovering from beam failures making it absolutely crucial for millimeter (mm) Wave  communication systems~\cite{key_challenges}. 

The BM framework in the current specifications includes beam sweeping, beam measurement and reporting, beam indication, beam failure detection and recovery~\cite{erik}. The beam measurement and reporting can be based on Synchronization Signal  (SS) blocks or channel state information RSs (CSI-RSs).
In D-MIMO, Reference Signal (RS) is transmitted in each DL beam by the APs, sequentially in time. Link quality (e.g., Layer $1$ Reference Signal Received Power (L1-RSRP)) on each DL RS is measured and reported by UE. 
Finding the best AP and beam (direction) requires measuring the DL channel from all the APs using all the beams and becomes resource-heavy, especially at mmWave frequencies due to a large number of narrow beams at the APs. Our objective is to investigate whether the best AP/beam can be inferred from sounding the DL channel on only a subset of APs and beams, with help of AI/ML. 

\begin{figure}[ht]
\begin{center}
  \includegraphics[scale=0.4]{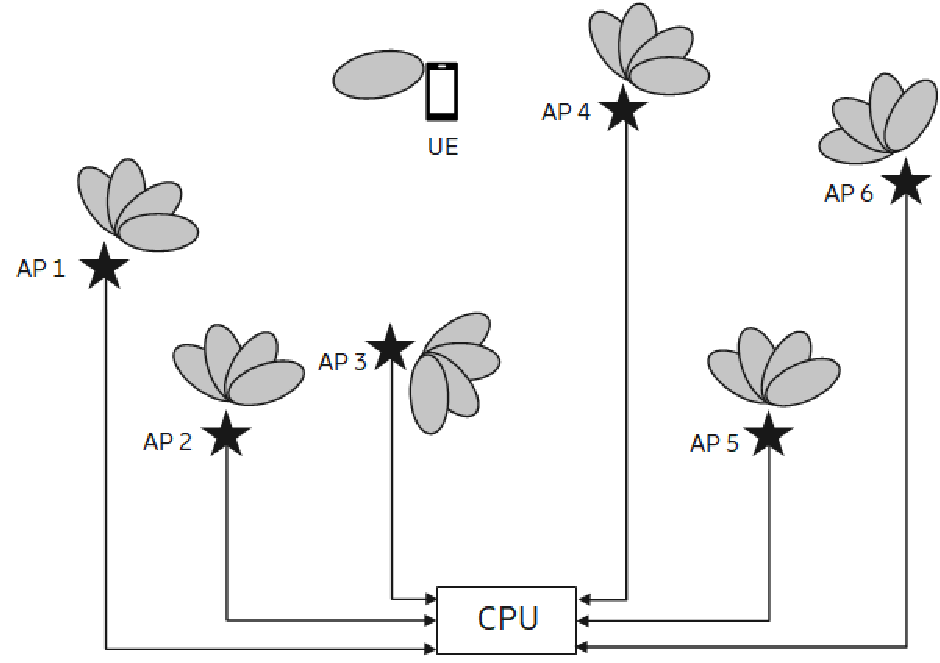} 
\end{center}
  \caption{Analog beamforming in mmWave D-MIMO}
  \label{fig:analog_bf}
\end{figure}
Fig.~\ref{fig:analog_bf} shows a D-MIMO with multiple APs ($AP_1$, $AP_2$,\ldots, $AP_6$) connected to a CPU and serving UE with analog beams. 
 Our main focus is on the beam measurement and reporting operation where the DL RS is sent only from a subset of beams from the APs and the UE provides the L1-RSRP for the scanned subset. This work is related to the 3GPP Release $18$ intra-cell spatial domain DL Transmit beam prediction. 


There are  $M$ APs and $N$ DL beams per AP resulting in a set $A$ containing a total of $MN$ beams. UEs measures on only a subset $B$ of $MN$ beams and reports the L1-RSRP measurements to the network/CPU. Our objective is to determine whether CPU can use only the available measurements to predict the L1-RSRPs of all un-measured/missing beams or 
top-$K$ beams/APs using state of the art ML algorithms. AI/ML algorithms can be used for
beam prediction in spatial and time domains to reduce
overhead and latency and improve beam selection accuracy.
We demonstrate the performance of a classical ML model Random Forest (RF) and its specialized version  MissForest (MF) for dealing with missing data. We also train a modern Deep Learning based conditional-Generative Adversarial Networks (c-GAN). We train these models for different cardinalities of the available measurements and show the gains obtained making them potential candidates for beam inference. 

The rest of the paper is organized as follows. Section~\ref{sec:inference} explains the beam inference and how the ML algorithms can potentially be applied. In  Section~\ref{sec:exp_approaches}, we describe RF, MF, and c-GAN algorithms with their performances being discussed in Section~\ref{sec:perf}. Section~\ref{sec:conclusion} concludes the work.
\section{Beam Inference and simple ML algorithms}
\label{sec:inference}
\begin{figure}[ht]
\begin{center}
  \includegraphics[scale=0.35]{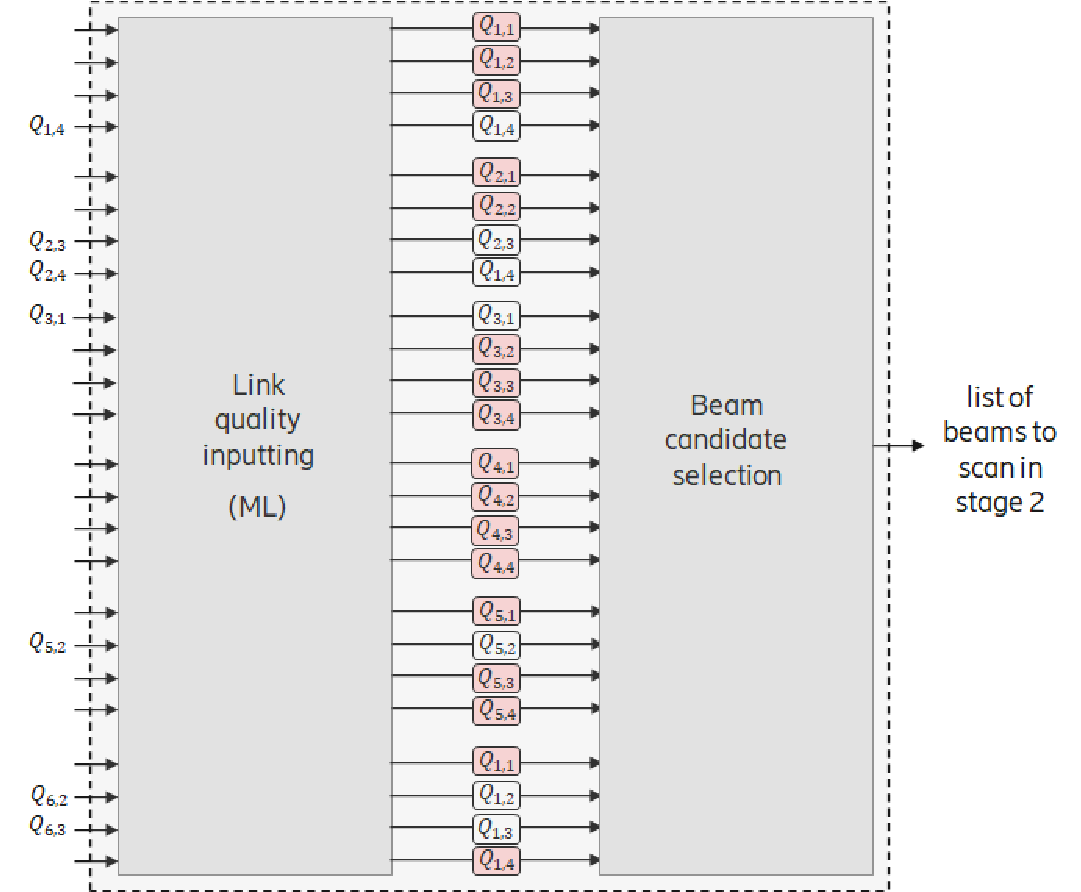} 
\end{center}
  \caption{Beam candidate inference}
  \label{fig:inference}
\end{figure}
\begin{figure}[ht]
\begin{center}
  \includegraphics[scale=0.4]{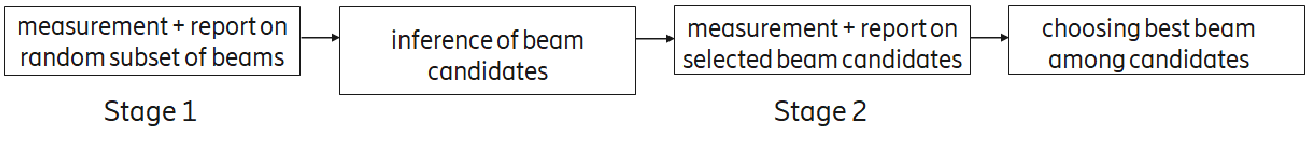} 
\end{center}
  \caption{Stages of Directed Search}
  \label{fig:stages}
\end{figure}

In the beam scanning framework, there can be two possible approaches, namely ``random search'', and ``directed search''. In the former, the UE provides the link quality report (L1-RSRP) for the scanned subset of beams. The directed search method consists of two phases, inference phase and the  beam selection phase for performing the directed scan. In Stage-1, we infer the link quailities that were not measured from the available measurements; using the beam candidate selection algorithm then chooses a list of beam candidates for a Stage-$2$ beam scan. One possible criterion is to scan $k$ (e.g. $k = 16$) beam candidates with highest quality metric $Q$ (from the mix of measured and inferred) and referred to as ``top-$k$'' method. The combined link quality and beam candidate inference is shown in Fig~\ref{fig:inference}. 
The inference algorithm inputs the list of beam candidates that are to be scanned where the DL RS is transmitted and the UE reports the link qualities of the beams. Once the reports are received, the beam with the best quality is chosen for future DL transmissions. The overall stages can be pictorially represented as in Fig~\ref{fig:stages}. 

The ML algorithm is trained using full measurements over time. In the later stage, the missing inputs are masked and the ML algorithm needs to learn to input missing values. Data is provided by UEs in different positions. A supervised learning algorithm (classifer or regressor) is used for infering the best beam. The data for training and testing was generated using a real $3D$ map and a ray tracing channel model. 
There are two possible methods for infering the beam candidates in ML  namely direct inference of beam candidates corresponding to classification and imputing missing measurements followed by beam selection corresponding to regression. 
\section{Explored Methodologies}
\label{sec:exp_approaches}
We explored both the supervised as well as the unsupervised approaches for performing Stage $2$ of the beam candidate inference wth different experimental approaches leading to varying performances. In the unsupervised method, we used the autoencoder (AE) framework~\cite{Goodfellow-et-al-2016}, where the network is trained on complete data and tested on incomplete data. The AE obtained compressed representation of the L1-RSRP values across UE positions and it was expected that they reproduce the patterns despite some values missing. However, the performance was unsatisfactory and is not presented here. 

In supervised learning, we can either obtain a direct inference (classification) or imputing + selection (regression). 
Missing Data is  quite prevalent in statistical analysis, and the imputation of missing values is a significant step in data analysis. 
In order to take care of missing values, there are a  lot of techniques, from simple mean/median/mode to more sophisticated methods. 
The classification performance is tested with FeedForward Neural Network (FFN), RF, MF,  and Generative Adversarial Networks (GANs). Training is executed on the complete data containing missing values and testing is on the incomplete one. With the 
imputing + selection approach, we demonstrate the exceptional performance of RF and MF  whereas the other algorithms perform poorly.

\subsection{Random Forest (RF)}
\label{sec:rf}
RF~\cite{Breiman2001} for data imputation is an exciting and efficient way of imputation, and it has almost every quality of being the best imputation technique. RFs are capable of scaling to significant data settings, and are robust to the non-linearity of data and efficient in handling outliers. RFs can handle both numerical and categorical data. On top of that, they have a built-in feature selection technique. These distinctive qualities of RFs can easily give it an upper hand over KNN or any other methods.

RFs are created from a large number of decision tree predictors, 
such that each tree is built by training on a random data set
sampled independently but from the same
distribution. 
After a large number of trees is generated, they vote for the most popular class.
RFs are based on the idea of Bagging~\cite{Breiman1996}, which involves perturbing and combining, i.e., building a model from a randomly sampled data set from the same distribution and then combining the models. The main effect of  bagging is to reduce variance or instability. 
The total error of a predictor can be expressed in terms of the bias error, variance error and irreducible error. Unstable classifiers such as decision trees characteristically have high variance and a low bias error. 
Hence, by applying Bagging on decision tree algorithms, the variance error is reduced. 

Tree-based models yield good predictions, with much less computational cost. It has been found in \cite{NEURIPS2022_0378c769} that tree based approaches like RFs and XGBoost outperfom Deep Learning methods on tabular data. Though Deep Learning methods have proven to perform very well on image, text data, their superiority in tabular data is unclear. 
RFs shows good performance on both real and categorical data, not requiring any scaling of the data which is important for Deep Learning methods.  


\subsection{Miss Forest (MF)}
\label{sec:mf}
MF~\cite{missforest} is arguably the best imputation algorithm to be used if precision is required and used as a benchmark for non-parametric imputation methods. 
It is implemented in $R$ language in the missForest() package. 
and the steps are:  

Step-1: Missing values are filled by the mean of respective columns for continuous, and mode for categorical data.

Step-2: Dataset is divided into two parts: training data consisting of the observed variables and the other is missing data used for prediction. Training and prediction sets are fed to RF, and the predicted data is imputed at appropriate places. After imputing all the values, one iteration gets completed.

Step-3: Step-2 is repeated until a stopping condition is reached. The iteration process ensures that the algorithm operates on better quality data in subsequent iterations. 

Step-4: Stopping condition can be either the sum of squared differences between the current and previous imputation increases or a specific iteration limit is reached. Usually, it takes $5-6$ iterations to attribute the data well.

In most common datasets and for different levels of missingness, MF outperformed other algorithms in some cases reducing the imputation error by more than $50\%$. The reason for the multiple iterations is that, from iteration 2 onwards, the RFs performing the imputation will be trained on better and better quality data that itself has been predictively imputed. 
\subsection{Generative Adversarial Networks (GANs)}
\label{ssection:beamgan}

\begin{figure}[!htbp]
      \includegraphics[scale=0.50, width=8.8cm]{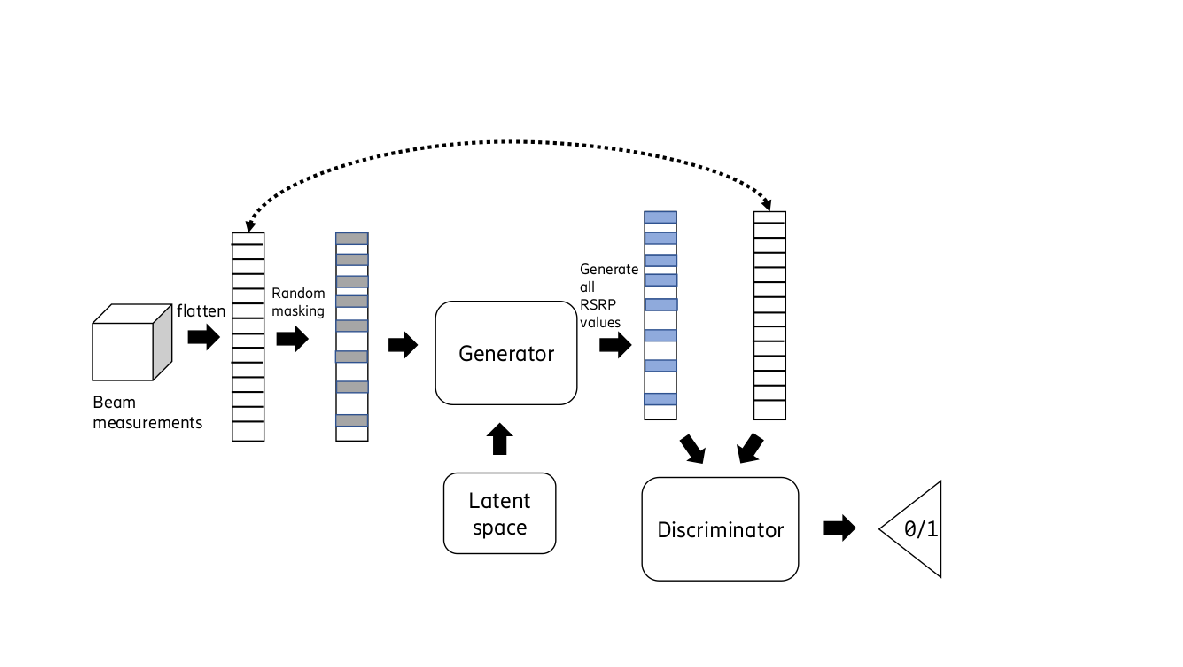} 
    \caption{Architecture of the proposed c-GAN. }
    \label{fig:beamganarch}    
\end{figure}

Generative adversarial networks (GANs) \cite{goodfellow2020generative} are a type of generative model which approximate the data generating distribution by generating realistic samples. GANs consist of a generator which takes random noise as input and produces realistic samples and a discriminator which learns to distinguish between generated and real samples. Although, they provide a very powerful modeling framework, GANs lack the ability to provide fine-grained control over the outputs. Conditional GANs (c-GANs) \cite{mirza2014conditional} have been proposed to endow GANs with a context variable providing more control over generated samples.  A c-GAN with a generator $G$ and a binary disciminator $D$ have the  objective function: 
$\min_G \max_D V(D, G) = E_{x \sim P_d} log D(x|c)  + E_{z \sim P_z}log(1 - D(G(z|c)|c))$, where $P_d$ denotes the distribution over the actual data, $P_z$ denotes the distribution of the generated data, $c$ is a context which acts as a conditional. Figure \ref{fig:beamganarch} provides an end-to-end view of the proposed c-GAN to fill out the masked beam measurement values. 
The masked measurement values are passed to the generator. It uses the unmasked values as context and transforms the latent space to fill the masked values. The disciminator then tries to predict whether the generated values are real values given the unmasked array of measurements. 
Both the generator and discriminator are fully connected feed-forward networks. The generator uses a latent dimension of 100 and has 4 linear layers with the first and last having 128 hidden units and the middle 2 having 256 hidden units. Each layer has a LeakyReLU activation. The output layer is a fully connected layer with 320 units. 
The discriminator has 3 hidden layers with 64 units each followed by a LeakyRelu activation. The generator is pre-trained in a self supervised way using the unmasked data for 50 epochs with an Adam optimizer with a learning rate of 1e-3 and a smooth L1 loss.  The full GAN is trained for 200 epochs with an Adam optimizer with a learning rate of 1e-5.

%

\begin{figure}[!htbp]
\begin{center}
  \includegraphics[scale=0.35]{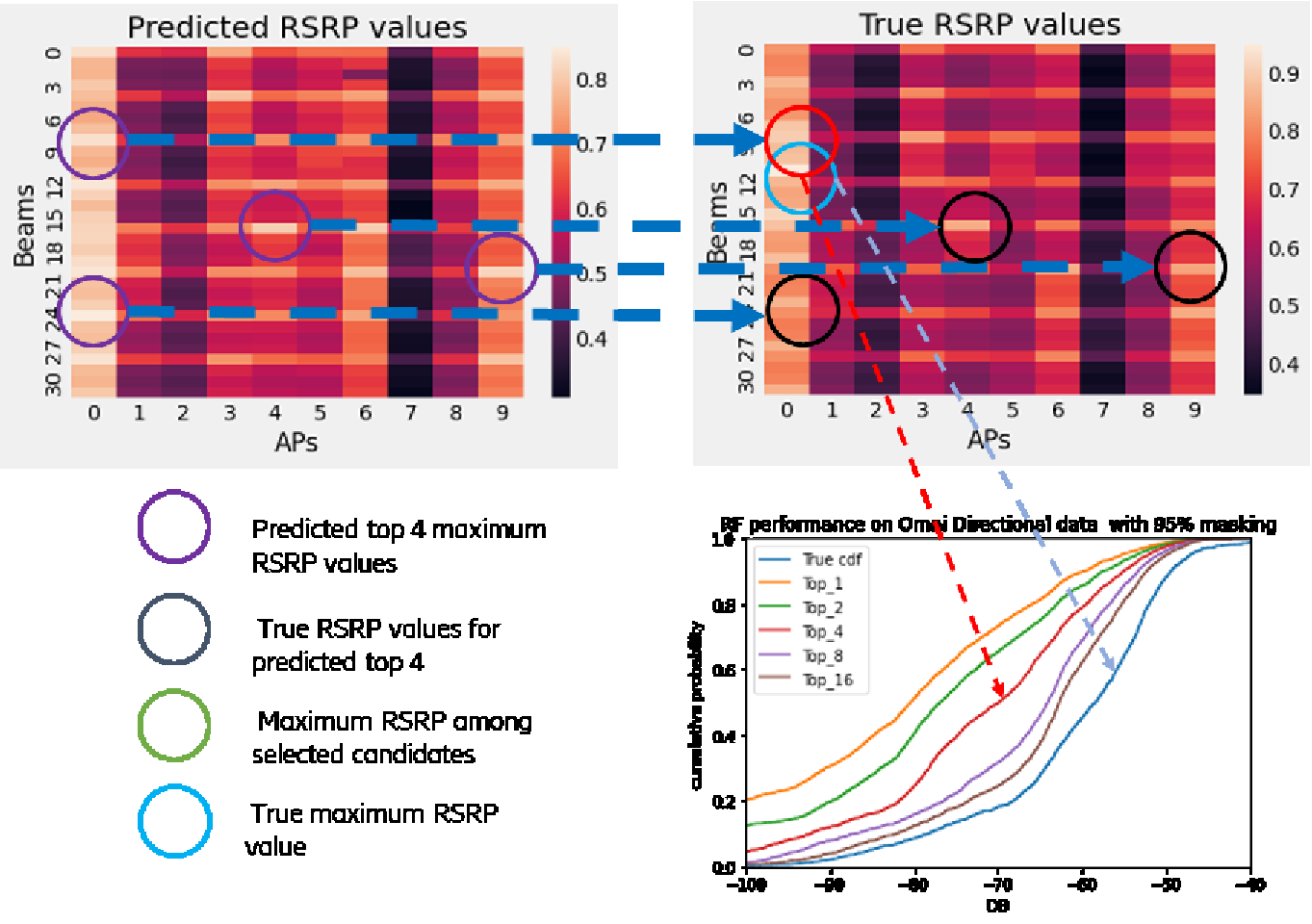} 
\end{center}
  \caption{\textbf{Comparison of CDF plots of L1-RSRP between top-$k$ predictions and ideal ones}}
  \label{fig:PE2}
\end{figure}

\begin{figure*}[ht]
\begin{center}
  \includegraphics[scale=0.35]{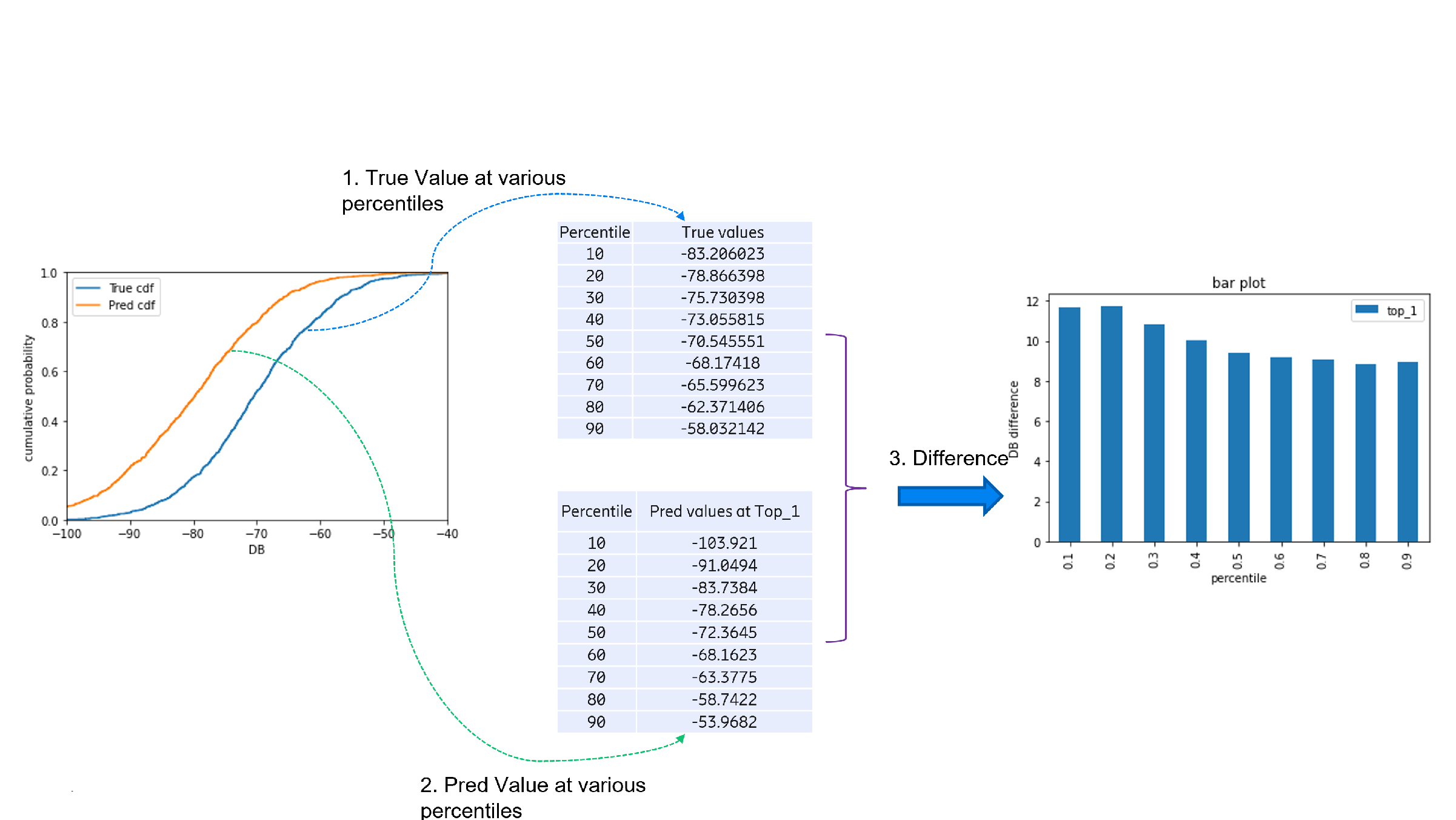} 
\end{center}
  \caption{\textbf{Relative performance at different percentiles: absolute difference in L1-RSRP values.
}}
  \label{fig:PE3}
\end{figure*}

\begin{figure}[!htbp]
\begin{center}
  \includegraphics[scale=0.8]{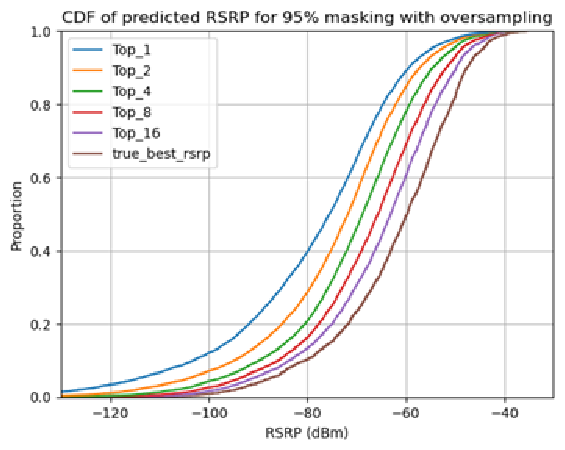} 
\end{center}
  \caption{CDF of predicted L1-RSRP with MF for $95\%$ masking random data with oversampling}
  \label{fig:cdf_impute_oversampling}
\end{figure}

\section{Performance Evaluation}
\label{sec:perf}
In this section, we explore the performance of RF and MF for data generated using a real $3D$ map with APs at fixed locations and UEs at different positions and a ray tracing channel model. Experiments were done on two datasets: one where the UEs are operating with an omnidirectional antenna, and the other where the UEs are equipped with a directional antenna with $9$ dB gain and are oriented randomly in horizontal plane. Different random orientations of the UEs during training and testing phase may make the inference of best AP/beam more challenging. It is of interest to quantify the performance penalty incurred by random UE orientation, as UEs will have directional antennas and be randomly oriented in practical deployments.

\subsection{Performance Evaluation measures}
\label{subsec:PEM}

One useful way of evaluating performance would be to compare the link quality metric (in our case, L1 - RSRP) of the actual best beam to link quality metric of the best beam from the set of best beams suggested by the ML algorithm (subset of beams in Stage 2 in Fig. 3). To clarify: the subset at Stage 2 may or may not contain the actual best beam. Therefore, the L1-RSRP of best beam among Stage 2 beams may or may not be equal to the L1 - RSRP of the actual best beam. If the RSRPs are different, it would be beneficial if the RSRP of best beam in Stage 2 is at least close to the RSRP of actual best beam. This would mean that there is only a small performance penalty in choosing the best beam suggested by ML as opposed to doing an exhaustive search.

We can visually judge a predictor's performance by comparing the cumulative
distribution function (CDF) of the true best L1-RSRP values with that of the L1-RSRP values of the best beam from Stage 2. Another way of presenting the same metric is to calculate the difference (in dB) between the true best L1-RSRP values with that of the L1-RSRP values of the best beam from Stage 2.

Figure~\ref{fig:PE2} illustrates the performance study with an example where the heat-maps have the different APs on the $X$-axis and the the different beam indices for each of the APs on the $Y$-axis. Lighter shades indicate higher L1-RSRP values.  The heat map on the left shows the predicted L1-RSRP values and the corresponding best indices.  
The heat map on the right shows the true L1-RSRP values, the maximum value and the true value at the predicted position. The CDF curves are for the true L1-RSRP values and the L1-RSRP values at the predicted indices and  smaller the distance between the curves, better is the performance.
Our goal is to obtain an acceptable performance with minimal readings and to predict the best index position or the best L1-RSRP value from the available options. If we predict the top-$k$ indices, obtain the corresponding readings and then use the best among these $k$ positions, we would desire that the estimated best value approaches the true best L1-RSRP value. This is illustrated in Figure~\ref{fig:PE2} with $k=4$. 
The heat-maps on the left and right show the predicted top $4$ positions/their values and the true best values for the predicted positions respectively. The CDF curves are for the true best L1-RSRP values and for the predicted top-$1, 2, 4, 8$ and $16$ beams. By plotting the CDF, we obtain an estimate of the deviation between the predicted and ideal L1-RSRP values. Increasing $k$ causes the CDF of predicted values to approach the CDF of the ideal L1-RSRP value.

The relative importance of the predictions for the L1-RSRP values is an important metric since accurate predictions are required only for the most prominent values. We show the absolute difference between the ideal L1-RSRP value and the best value from the top-$k$ predicted beams at various percentiles in Figure~\ref{fig:PE3} to demonstrate the relative performance. We provide additional results/discussions in the next section.


\subsection{Model performance}
We seek to find the minimum number of readings required to identify the index with minimum absolute difference between predicted L1-RSRP and the ideal value for a given UE position.
Each L1-RSRP value to be measured has an associated time/feedback  that should be minimized. RF and MF use $x$ measurements and obtain estimates of the top-$k$ candidates. Predictions for the $k$ candidates are combined with the available $(x$ +$k)$ measurements and the best beam is chosen. Our objective is to determine the best combination of $(x, k)$ that gives acceptable estimates of the best beam and its quality level. So, we mask a certain percentage of the available data, use the remaining unmasked data along with the unmasked data  
and study  the performance of RF and MF models with different masking percentages. 

In order to generate additional data with masking at different beams in each UE position, we oversample the original data by a certain amount for training the RF and MF models. 
With oversampling = $10$, each row in the data is repeated $10$ times with 
masking at different random points on each row but with the same masking percentage. The goal is to enlarge the dataset and obtain more data for training and test and to improve the insights. Fig~\ref{fig:cdf_impute_oversampling} shows the CDF plot with oversampling and $95\%$ masking for different top $k$ candidates. It is clear that as $k$ increases, the respective CDF  approaches the CDF  of the ideal L1-RSRP. 

\begin{figure*}[ht]
    \begin{center}
      \includegraphics[scale=0.85]{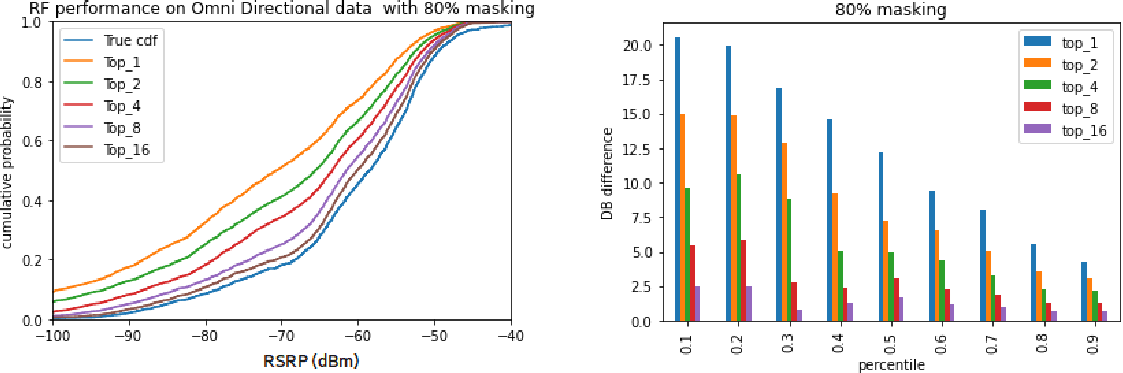} 
    \end{center}
      \caption{CDF of predicted L1-RSRP with RF for $80\%$ masking normal data}
      \label{fig:rf_80_normal}
    \end{figure*}

\begin{figure*}[ht]
\begin{center}
  \includegraphics[scale=0.9]{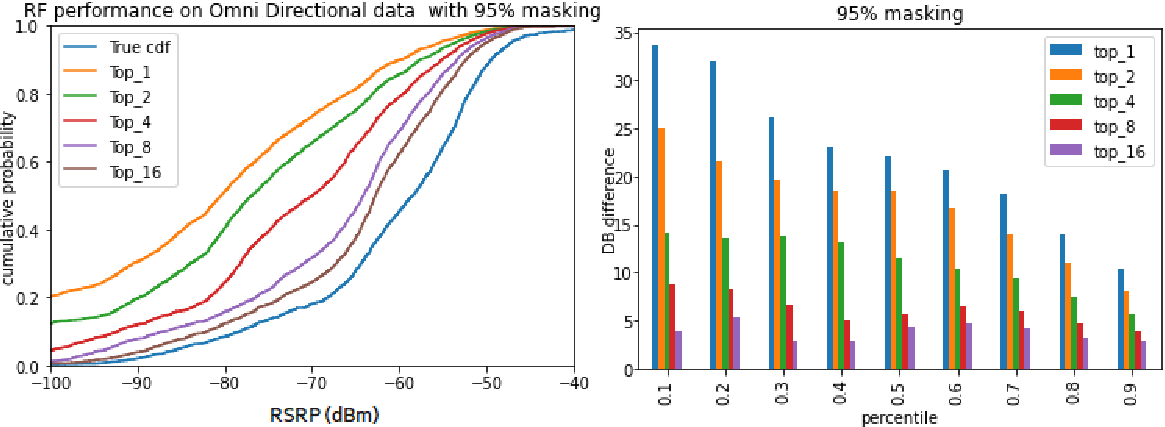} 
\end{center}
  \caption{CDF of predicted L1-RSRP with RF for normal data}
  \label{fig:rf_95_normal}
\end{figure*}

\begin{figure*}[ht]
\begin{center}
  \includegraphics[scale=0.9]{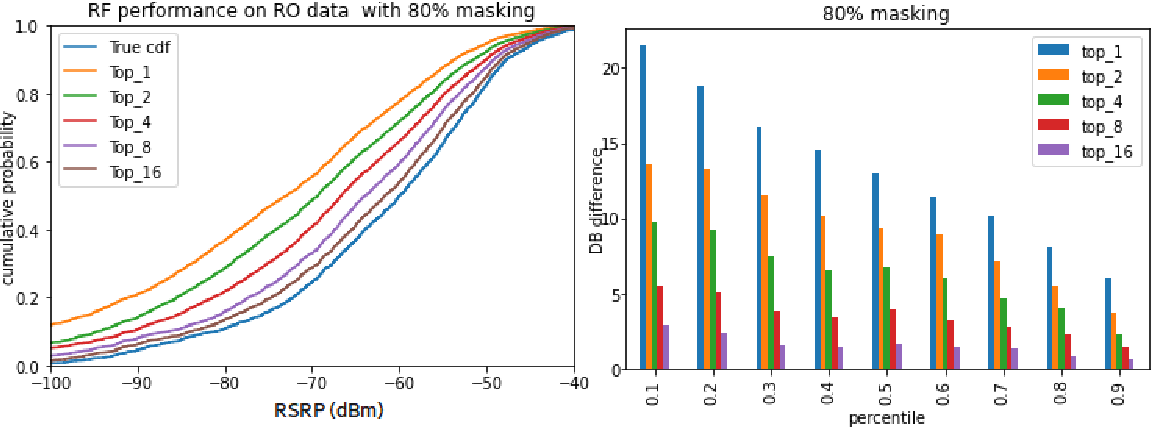} 
\end{center}
  \caption{CDF of predicted L1-RSRP with RF for random data}
  \label{fig:rf_80_random}
\end{figure*}

\begin{figure*}[ht]
\begin{center}
  \includegraphics[scale=0.9]{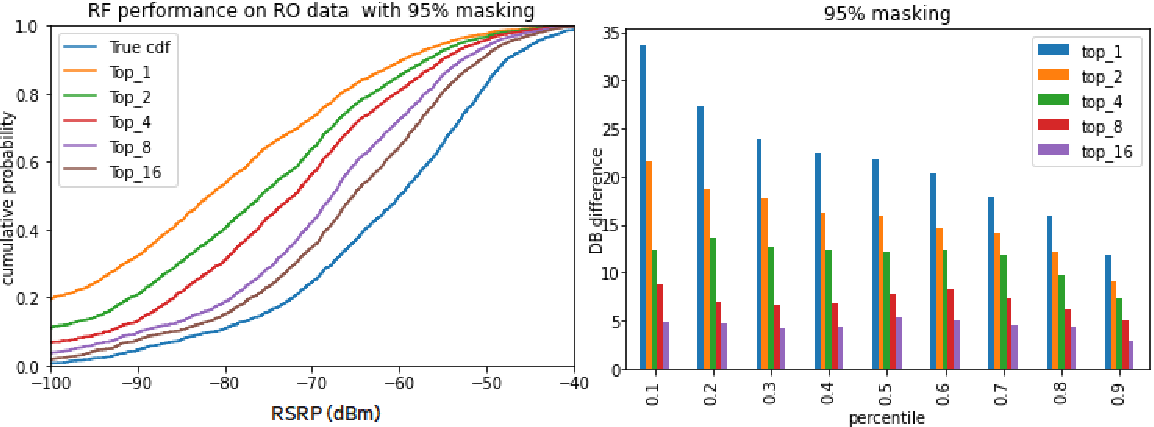} 
\end{center}
  \caption{CDF of predicted L1-RSRP with RF for random data}
  \label{fig:rf_95_random}
\end{figure*}

\begin{figure*}[ht]
\begin{center}
  \includegraphics[scale=0.9]{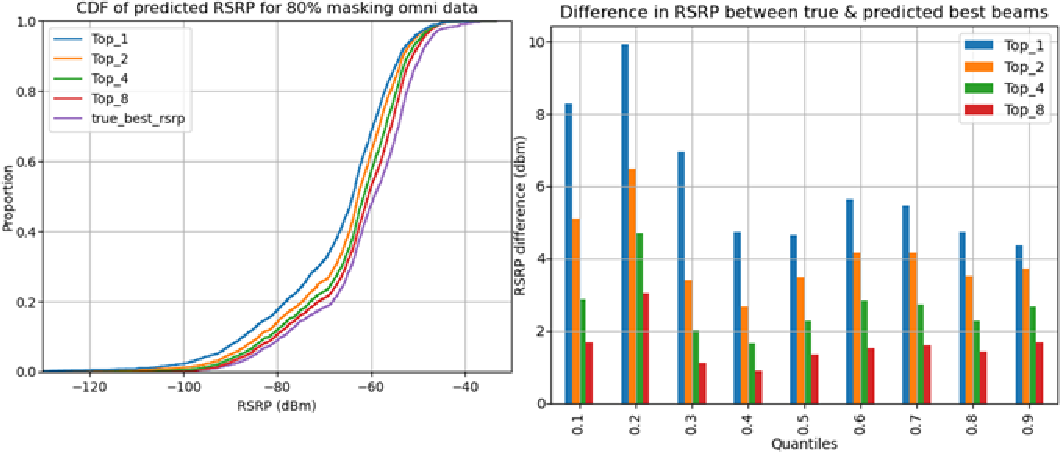} 
\end{center}
  \caption{CDF of predicted L1-RSRP with MF for $80\%$ masking normal data}
  \label{fig:cdf_80_normal}
\end{figure*}

\begin{figure*}[ht]
\begin{center}
  \includegraphics[scale=0.8]{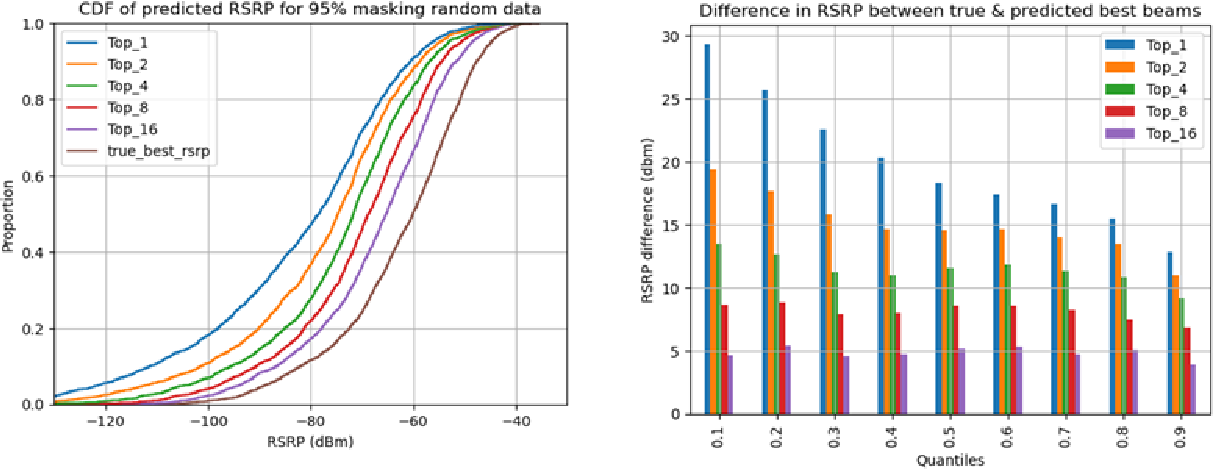} 
\end{center}
  \caption{CDF of predicted L1-RSRP with MF for $95\%$ masking random data}
  \label{fig:cdf_95_random}
\end{figure*}

Figures~\ref{fig:rf_80_normal} -~\ref{fig:cdf_95_random} provide the CDF plots for different masking percentages with RF and MF models. In addition, the plots also show the absolute difference in L1-RSRP values between the true best beam and top-$k$, $k=1, 2, 4, 8, 16$ predicted candidates. We could observe that the CDF plots of the top-$k$ candidates approach the true CDF as $k$ increases and the absolute difference between predicted and ideal value stays below $5$ dB in all the cases.  The results from our c-GAN are shown in Figure \ref{fig:ganresults}. The deviations from the c-GAN are slightly larger than both RF and MF, but the GAN is expected to scale better with data size and novel model architectures. 

To compare the distribution of the true best RSRP from the 3 distributions generated by each of our proposed approaches we use the 1-Wasserstein(1-W) distance, originating from Optimal Transport theory \cite{wasserstein}.  Intuitively,  given 2 probability distributions, the 1-W distance measures the work required to transform one distribution into the other. It overcomes the drawbacks of entropy based measures like the Kulback-Leibler(KL) and Jensen-Shannon(JS) divergence; it also accounts for the geometry of the underlying space which is an important consideration for comparing probability distributions. 
We report our results in Table \ref{tab:comparew1}. The RF and MF outperfom the c-GAN consistently for all top-K beams selected. However, the pairwise differences between RF and MF are smaller than those with RF/MF and c-GAN.

\begin{figure*}[ht]
\begin{center}
  \includegraphics[scale=0.9]{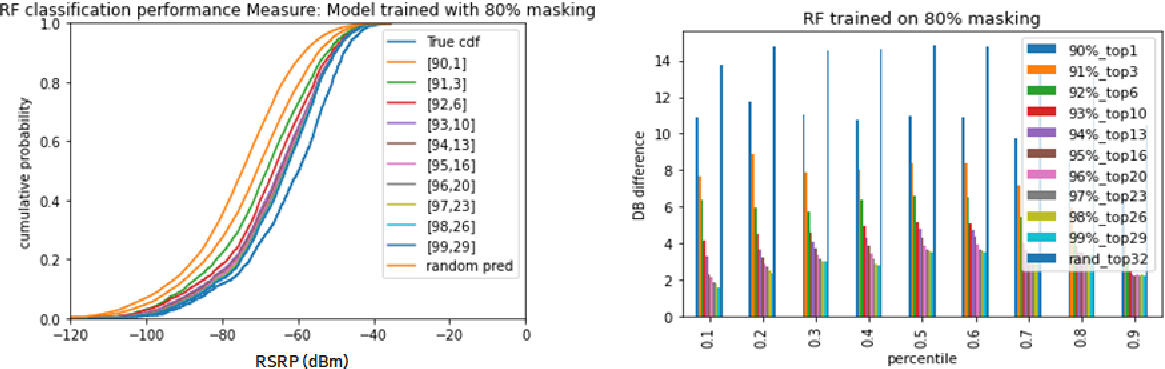} 
\end{center}
  \caption{CDF of predicted L1-RSRP with RF with different masking on test data}
  \label{fig:rf_80_diff_masking}
\end{figure*}

\begin{figure*}[ht]
    \includegraphics[scale=0.5, clip]{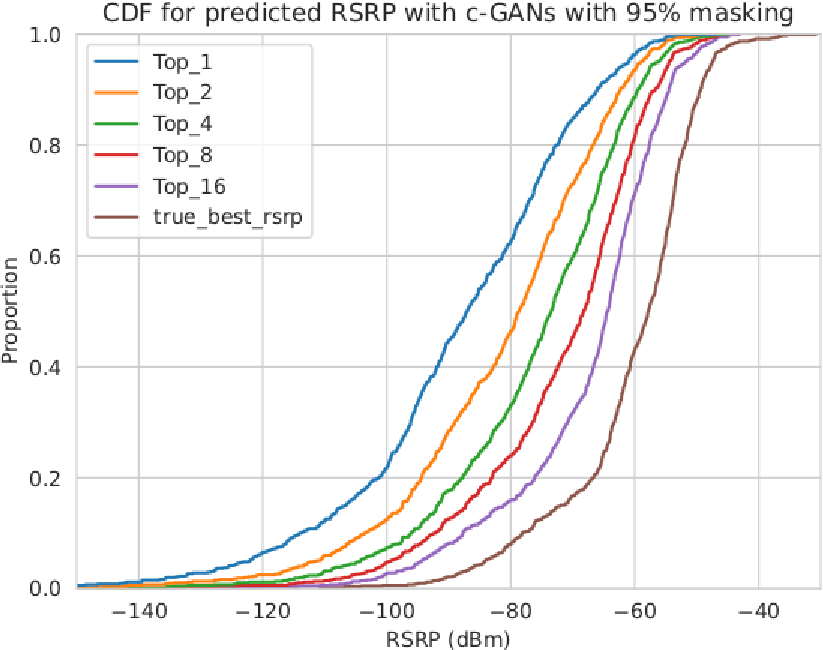} \includegraphics[scale=0.5, clip]{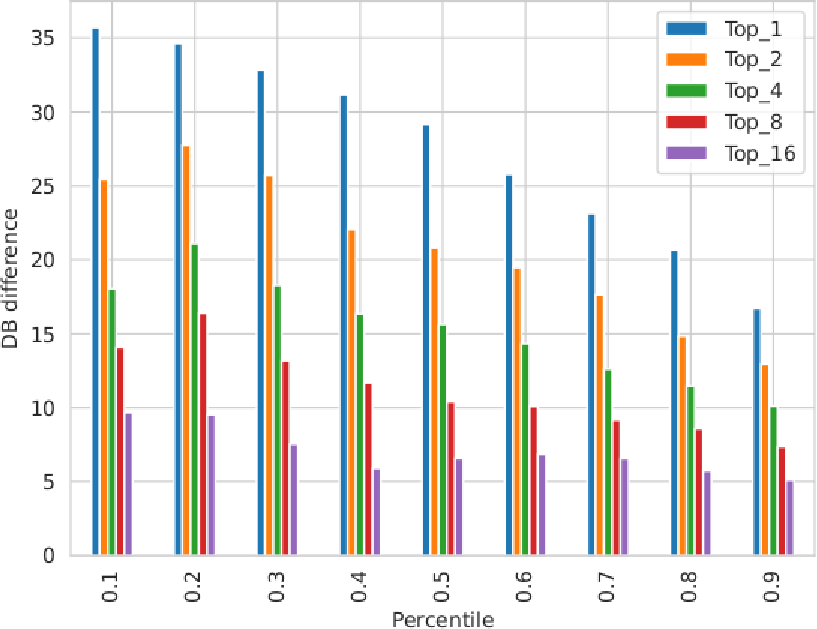}
    \caption{CDF of predicted L1-RSRP with c-GANs with 95\% masked data }
    \label{fig:ganresults}    
\end{figure*}

An additional aspect is to determine the kind of data to train the RF and MF models for yielding the best results, \emph{i.e.}, what should be the percentage of masked data on which the models should be trained to give decent performance while retaining robustness to different masking on test data. Figure~\ref{fig:rf_80_diff_masking} provides results when RF model is trained on $80\%$ masked data but tested with varying masking percentages and different top-$k$ beams. For instance, with $93\%$ masking, the estimates of the top-$10$ beams are used for plotting the CDF and absolute differences. A random sampling with top-$32$ beams is also shown in the plot demonstrating that the performance gap reduces with higher values of $k$.

\section{Conclusion}
We studied the beam prediction in the beam management framework for overhead and latency reduction where measurements are available only for a subset of the beams from the APs, and demonstrated high beam prediction accuracy with Random Forest and MissForest algorithms. We also trained a Deep Generative adversarial network, whose performance was affected by the limited data available. However, we expect the model to have superior performance with a larger dataset size. The study can be extended where we can build an initial model that can be dynamically updated based on the environment. The positions of the APs can be used as a priori information by the UE to determine the beam scanning activity. In practical scenarios, beams with azimuth angles pointing towards the horizon will occur more often than beams with other azimuth. It will be interesting to explore the ability of RF/MF/GAN models to predict the uncommon beams and to visualize their performance on such rare events. 
We can also design an explainable model for understanding the prediction of L1-RSRP values and best beams or a causal model for determining the beam scan. 
\label{sec:conclusion}

\begin{table}[ht]
  \centering
   \begin{tabular}{||c | c | c | c||} 
   \hline
   top-K & Random Forest(RF) & MissForest(MF) & c-GAN \\ [0.5ex] 
   \hline\hline
   1 & 13.02 & 17.74 & 27.70 \\ 
   2 & 9.76 & 12.94 & 20.88 \\
   4 & 6.89 & 9.72 & 15.58 \\
   8 & 4.14 & 6.82 & 11.38 \\
   16 & 3.12 & 4.32 & 7.13 \\ [1ex] 
   \hline
   \end{tabular}
   \caption{Comparison of 1-Wasserstein(1-W) distances from the true best RSRP distribution with 95\% masking applied for different top-K values.}
   \label{tab:comparew1}
  \end{table}

\bibliographystyle{IEEEtran}
\bibliography{IEEEabrv, reference1.bib}

\begin{thebibliography}{10}
\providecommand{\url}[1]{#1}
\csname url@samestyle\endcsname
\providecommand{\newblock}{\relax}
\providecommand{\bibinfo}[2]{#2}
\providecommand{\BIBentrySTDinterwordspacing}{\spaceskip=0pt\relax}
\providecommand{\BIBentryALTinterwordstretchfactor}{4}
\providecommand{\BIBentryALTinterwordspacing}{\spaceskip=\fontdimen2\font plus
\BIBentryALTinterwordstretchfactor\fontdimen3\font minus
  \fontdimen4\font\relax}
\providecommand{\BIBforeignlanguage}[2]{{%
\expandafter\ifx\csname l@#1\endcsname\relax
\typeout{** WARNING: IEEEtran.bst: No hyphenation pattern has been}%
\typeout{** loaded for the language `#1'. Using the pattern for}%
\typeout{** the default language instead.}%
\else
\language=\csname l@#1\endcsname
\fi
#2}}
\providecommand{\BIBdecl}{\relax}
\BIBdecl

\bibitem{key_challenges}
Y.~Heng, J.~G. Andrews, J.~Mo, V.~Va, A.~Ali, B.~L. Ng, and J.~C. Zhang, ``Six
  key challenges for beam management in 5.5g and 6g systems,'' \emph{IEEE
  Communications Magazine}, vol.~59, no.~7, pp. 74--79, 2021.

\bibitem{erik}
E.~Dahlman, S.~Parkvall, and J.~Skold, \emph{5G NR: The Next Generation
  Wireless Access Technology}, 1st~ed.\hskip 1em plus 0.5em minus 0.4em\relax
  USA: Academic Press, Inc., 2018.

\bibitem{Goodfellow-et-al-2016}
I.~Goodfellow, Y.~Bengio, and A.~Courville, \emph{Deep Learning}.\hskip 1em
  plus 0.5em minus 0.4em\relax MIT Press, 2016,
  \url{http://www.deeplearningbook.org}.

\bibitem{Breiman2001}
\BIBentryALTinterwordspacing
L.~Breiman, ``{Random Forests},'' \emph{Machine Learning}, vol.~45, pp. 5--32,
  10 2001. [Online]. Available: \url{https://doi.org/10.1023/A:1010933404324}
\BIBentrySTDinterwordspacing

\bibitem{Breiman1996}
\BIBentryALTinterwordspacing
------, ``{Bagging predictors},'' \emph{Machine Learning}, vol.~24, pp.
  123--140, 8 1996. [Online]. Available:
  \url{https://doi.org/10.1007/BF00058655}
\BIBentrySTDinterwordspacing

\bibitem{NEURIPS2022_0378c769}
L.~Grinsztajn, E.~Oyallon, and G.~Varoquaux, ``Why do tree-based models still
  outperform deep learning on typical tabular data?'' in \emph{Advances in
  Neural Information Processing Systems}, S.~Koyejo, S.~Mohamed, A.~Agarwal,
  D.~Belgrave, K.~Cho, and A.~Oh, Eds., vol.~35.\hskip 1em plus 0.5em minus
  0.4em\relax Curran Associates, Inc., 2022, pp. 507--520.

\bibitem{missforest}
\BIBentryALTinterwordspacing
D.~J. Stekhoven and P.~Bühlmann, ``{MissForest—non-parametric missing value
  imputation for mixed-type data},'' \emph{Bioinformatics}, vol.~28, no.~1, pp.
  112--118, 10 2011. [Online]. Available:
  \url{https://doi.org/10.1093/bioinformatics/btr597}
\BIBentrySTDinterwordspacing

\bibitem{goodfellow2020generative}
I.~Goodfellow, J.~Pouget-Abadie, M.~Mirza, B.~Xu, D.~Warde-Farley, S.~Ozair,
  A.~Courville, and Y.~Bengio, ``Generative adversarial networks,''
  \emph{Communications of the ACM}, vol.~63, no.~11, pp. 139--144, 2020.

\bibitem{mirza2014conditional}
M.~Mirza and S.~Osindero, ``Conditional generative adversarial nets,'' 2014.

\bibitem{wasserstein}
S.~Vallender, ``Calculation of the \uppercase{W}asserstein distance between
  probability distributions on the line,'' \emph{Theory of Probability \& Its
  Applications}, vol.~18, no.~4, pp. 784--786, 1974.

\end{thebibliography}


\end{document}